# A Longitudinal Higher-Order Diagnostic Classification Model

Peida Zhan (Zhejiang Normal University)[1]

Hong Jiao (University of Maryland, College Park)

Dandan Liao (American Institutes for Research)

**Note:**

This is an arXiv preprint, may not be the final version [Manuscript submitted for publication and currently under review]. For reference:

Zhan, P., Jiao, H., & Liao, D. (2017). A longitudinal higher-order diagnostic classification model. *arXiv preprint 1709.03431.* URL https://arxiv.org/abs/1709.03431

---

[1] Corresponding author: Peida Zhan, Department of Psychology, College of Teacher Education, Zhejiang Normal University, No. 688 Yingbin Road, Jinhua, Zhejiang, 321004, P. R. China. Email: pdzhan@gmail.com

## A Longitudinal Higher-Order Diagnostic Classification Model

## Abstract

Providing diagnostic feedback about growth is crucial to formative decisions such as targeted remedial instructions or interventions. This paper proposed a longitudinal higher-order diagnostic classification modeling approach for measuring growth. The new modeling approach is able to provide quantitative values of overall and individual growth by constructing a multidimensional higher-order latent structure to take into account the correlations among multiple latent attributes that are examined across different occasions. In addition, potential local item dependence among anchor (or repeated) items can also be taken into account. Model parameter estimation is explored in a simulation study. An empirical example is analyzed to illustrate the applications and advantages of the proposed modeling approach.

***Keywords*** cognitive diagnosis; diagnostic classification model; longitudinal data; anchor-item; local item dependence; DINA

## A Longitudinal Higher-Order Diagnostic Classification Model

## Introduction

The central topic in educational research and assessment is to measure change in student learning on different occasions (Fischer, 1995). Measuring individual growth or change relies on longitudinal data collected over multiple measures of achievement construct along the growth trajectory (Wang, Jiao, & Zhang, 2013; Wang, Kohli, & Henn, 2016). Up to now, several researches concerning individual or overall changes have been conducted in fields such as developmental, educational and applied psychology.

In recent years, cognitive diagnosis has received great attention, particularly in the areas of educational and psychological measurement (Rupp, Templin, & Henson, 2010). One of the main objectives of cognitive diagnosis is to evaluate respondents' status of mastery or non-mastery of skills (also called "attributes") and then provides diagnostic feedback for teachers or clinicians to help them make decisions regarding remedial teachings or targeted interventions (Zhan, Jiao, & Liao, 2018). Several diagnostic classification models (DCMs), also known as cognitive diagnosis models, have been developed, such as the deterministic-inputs, noisy "and" gate (DINA) model (Haertel, 1989; Junker & Sijtsma, 2001; Macready & Dayton, 1977) and the deterministic-inputs, noisy "or" gate (DINO) model (Templin & Henson, 2006). Some general DCMs are also available (de la Torre, 2011; Henson, Templin, & Willse, 2009; von Davier, 2008). However, most DCMs do not concern about measuring growth in terms of several possibly related attributes over multiple occasions, which could be potentially very important for remedial teaching or targeted intervention.

Unlike continuous latent variables in the item response theory (IRT) models, the attributes in DCMs are categorical (typically, binary). Therefore, the methods for modeling growth in the IRT framework may not be directly extended to capture growth in the mastery of attributes. For example, the change in the mastery of attributes may not be directly modeled by the variance-covariance-based methods (Collins & Wugalter, 1992) when assuming multiple continuous latent variables follow a multivariate normal distribution (e.g., Andrade & Tavares, 2005; von Davier, Xu, & Carstensen, 2011).

In DCMs, to account for change in attributes, Li, Cohen, Bottge, and Templin (2015) proposed a latent transition analysis (LTA; Collins & Wugalter, 1992), also known as mixed hidden (or latent) Markov model (Van de Pol & Langeheine, 1990), in combination with the DINA model in repeated measures. Likewise, Kaya and Leita (2017) combined the LTA with the DINA model and the DINO model, respectively. Such LTA-based methods provided an attribute-level transition probability matrix rather than a quantitative value of change which was used more commonly. Additionally, it assumed that attributes are independent and their transition probabilities are also independent. However, those independence assumptions may be tenuous as the attributes may be correlated (de la Torre & Douglas, 2004; Rupp et al., 2010). Recently, focusing on modeling learning trajectory, Wang, Yang, Culpepper, and Douglas (2017) proposed a higher-order, hidden Markov model for attribute transitions. Compared with above two LTA-based methods, Wang et al.'s model used a set of observed and latent covariates, such as intervention indicators and a time-invariant general learning ability, to model the attribute-level transition probabilities. The correlations among attributes on the first occasion and the correlations among different transition probabilities were also accounted for. Additionally, Wang et al.'s model assumed learning trajectories to be non-decreasing. Rather than employing attribute-level hidden Markov models, Chen, Culpepper, Wang, and Douglas (2017) considered an attribute-pattern-level approach for approximating the learning trajectory space. In Chen et

al.'s model, the attribute-pattern-level transition probability matrix explicitly provides the probabilities of remaining in the same pattern or changing to other patterns from one occasion to the next one. However, Chen et al.'s model assumed the transition probabilities of different attribute patterns were the same on different occasions, which were also the same for each individual.

Essentially, these transition probability-based methods analyzed the longitudinal data from the latent class modeling perspective, which can all be taken as a special case or an application of the mixture hidden Markov model (Vermunt, Tran, & Magidson, 2008). Moreover, despite the use of the repeated measures design in these studies, local item dependence among a person's responses to the repeated items on different occasions (Cai, 2010; Paek, Park, Cai, & Chi, 2014) was not taken into account. In the IRT framework, it has been demonstrated that the local item dependence affects model parameter estimation, equating, and estimation of test reliability (e.g., Bradlow, Wainer, & Wang, 1999; Jiao, Kamata, Wang, & Jin, 2012; Jiao & Zhang, 2015; Sireci, Tissen, & Wainer, 1991; Tao & Chao, 2016; Wang & Wilson, 2005; Zhan, Wang, Wang, & Li, 2014). Similarly, in the DCM framework, if local item dependence is ignored, large estimation errors of item parameters could appear, and the correct classification rate of attributes might reduce (Zhan, Li, Wang, Bian, & Wang, 2015; Zhan, Liao, & Bian, 2018).

Additionally, Hansen (2013) proposed a longitudinal unidimensional DCM for the repeated measures when only one attribute is required on each occasion, and multiple attributes are required on different occasions. Further, a higher-order latent structural model (also known as the higher-order latent trait model) (de la Torre & Douglas, 2004) is employed to account for associations among the attributes across different time points, where local item dependence among repeated items were accounted for by using additional random-effect latent variables.

Although theoretically DCMs have been employed to measure multiple dimensions of latent constructs rather than a unidimensional attribute on a given occasion, this model provides insight of longitudinal analysis in diagnostic assessment from another perspective that is different from transition probability-based methods.

This study proposes a new longitudinal diagnostic classification modeling approach for measuring growth, which can be used in not only the repeated measures design but also the anchor-item design. Among numerous DCMs, the interpretability of the DINA model makes it be the most popular one. Thus, in this study, the DINA model is taken as an example to illustrate the conceptualization of the proposed modeling approach. The proposed method can be easily extended to many other DCMs, such as the log-linear DCM (LCDM) (Henson et al., 2009) and its special cases. The rest of the paper starts with a review of the DINA model with a higher-order latent structure. Then the proposed longitudinal DINA (denoted as the *Long-DINA*) model is presented and illustrated. Item response data from a physical achievement test was analyzed to illustrate the application of the proposed modeling approach.

## Longitudinal Diagnostic Classification Modeling

### DINA Model with a Higher-Order Latent Structure

Let $Y_{ni}$ be the observed response of person $n$ to item $i$. In the DINA model, the relationship among attributes and an observed response can be expressed as (DeCarlo, 2011; Rupp et al., 2010; von Davier, 2014)

$$\text{logit}(P(Y_{ni} = 1 \mid \boldsymbol{\alpha}_n)) = \lambda_{i0} + \lambda_{i(K)} \prod_{k=1}^{K} \alpha_{nk}^{q_{ik}} , \qquad (1)$$

where $\text{logit}(x) = \log(x / (1-x))$; $P(Y_{ni} = 1 \mid \boldsymbol{\alpha}_n)$ is the probability of a correct response by person $n$ to item $i$; $\lambda_{i0}$ and $\lambda_{i(K)}$ are the intercept and the $K$-way interaction effect parameters, respectively, for item $i$. In such a case, the guessing and slipping probability of item $i$ ($g_i$ and $s_i$) can be expressed respectively as follows:

$$g_i = \frac{\exp(\lambda_{i0})}{1 + \exp(\lambda_{i0})} \quad \text{and} \quad s_i = 1 - \frac{\exp(\lambda_{i0} + \lambda_{i(K)})}{1 + \exp(\lambda_{i0} + \lambda_{i(K)})};$$

$\alpha_{nk}$ is the attribute for person $n$ on attribute $k$ ($k = 1,..., K$), with $\alpha_{nk} = 1$ if person $n$ masters attribute $k$, and $\alpha_{nk} = 0$ otherwise. Q-matrix (Tatsuoka, 1983) is a $I \times K$ matrix with element $q_{ik}$ indicating whether attribute $k$ is required to answer item $i$ correctly; $q_{ik} = 1$ if the attribute is required, and 0 otherwise.

In practice, attributes in a test are often correlated. In such cases, it may be assumed that a general continuous latent ability underlies these attributes. Let $\alpha_{nk}$ be person $n$'s attribute $k$ and $\theta_n$ be the general ability of person $n$. The probability of $\alpha_{nk} = 1$ conditional on $\theta_n$ is defined as follows (de la Torre & Douglas, 2004),

$$\text{logit}(P(\alpha_{nk} = 1 \mid \theta_n)) = \delta_k \theta_n + \beta_k, \tag{2}$$

where $\delta_k$ and $\beta_k$ are the slope and intercept parameters of attribute $k$, respectively. To reduce computational burden, the attribute slope parameter $\delta_k$ can be further constrained as $\delta_k = \delta$, suggesting all attributes share the same slope parameter (de la Torre, Hong, & Deng, 2010), or $\delta_k$ = 1 for all attributes (Ma & de la Torre, 2016), similar to the Rasch modeling.

## Modeling Growth in DCMs

***Basic modeling***. In DCMs, attributes are typically modeled as categorical, especially binary variables. Thus, the longitudinal modeling approaches within the IRT framework such as the multivariate normal distribution strategy (e.g., von Davier et al., 2011) and the latent growth (curve) model-based strategy (e.g., Wang et al., 2016) cannot be employed directly. However, the general continuous latent trait $\theta$ in the higher-order latent structural model (Equation 2) can be an alternative.

The proposed model for two-time points can be graphically presented in Figure 1; it can also be extended to more time points. The DINA model (or other DCMs) is specified as the first-order model to link the attributes of a respondent to the observed response data at each time point. Further, a second-order latent structural model is specified to determine the mastery status for attributes of the respondents. Thus, at a given time point, the first two orders represent the higher-order DINA model (Equation 2). For the proposed model, the relationship between the general latent traits measured at different time points is specified at the third order. In other words, the Long-DINA model is a multidimensional extension of the higher-order DINA model, but the multidimensionality does not refer to different general ability dimensions rather the same general ability measured at different time points. Theoretically, this third-order model utilizes both strategies for measuring individual growth, i.e., the multivariate normal distribution strategy and the latent growth model-based strategy. As the repeated measures design is not always feasible in educational measurement, a more common practice of test administration over time involves multiple test forms that share anchor-items. This design is called the anchor-item design, such as the nonequivalent groups with anchor test (NEAT) design; however, it may induce local item dependence among a respondent's responses to the same anchor-items on multiple occasions. Therefore, additional random-effects latent variables or testlet-effects (Bradlow et al., 1999; e.g., $\gamma_1$ in Figure 1) can be introduced to account for local item dependence. The number of such random-effects variables is the same as the number of anchor items (Cai, 2010).

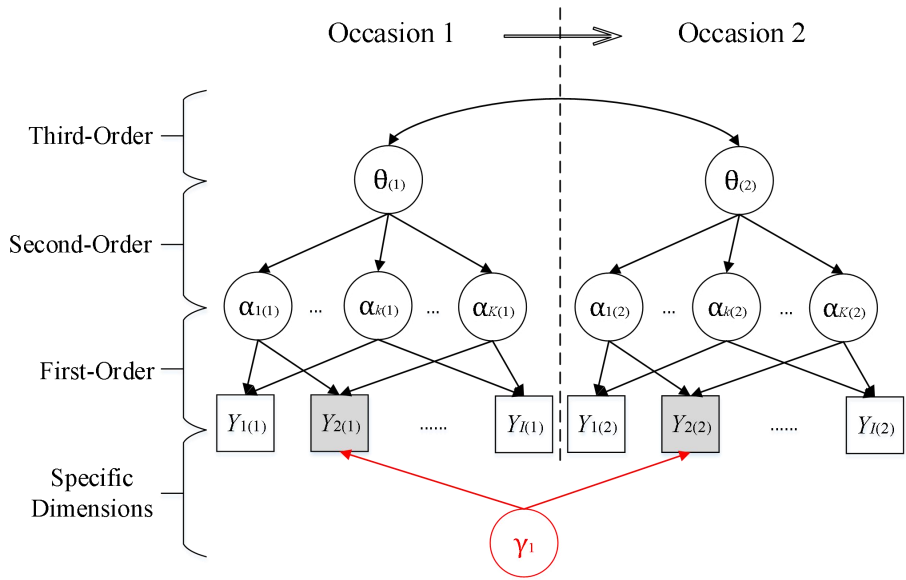

**Figure 1**. A graphical representation of the Long-DINA model for two-time points.

***First-order***. In Figure 1, responses $Y_{2(1)}$ and $Y_{2(2)}$ are for the same anchor-item at two time points. The specific factor $\gamma_1$ should be added in the first-order model to capture local item dependence. To account for local item dependence in DCMs, Hansen (2013) and Hansen, Cai, Monroe, and Li (2016) proposed a higher-order, hierarchical DCM which can be viewed as a combination of the two-tier item factor model (Cai, 2010) and the LCDM. Like the two-tier item factor model, Hansen's model can only account for local item dependence due to one source. Zhan et al. (2015) proposed (within-item) multidimensional testlet-effect DCMs which simultaneously account for multiple sources of local item dependence within one item (Rijmen, 2011; Zhan et al., 2014). Multiple within-item local item dependence may be presented in assessment when testlet-based items are repeatedly used or used as anchor-items (e.g., Zu & Liu, 2010). However, modeling an additional tier of specific factors could substantially increase model complexity. To simplify the proposed model, only one source of local item dependence is modeled in this study.

Following Hansen's and Zhan et al.'s models, for a given occasion $t$ ($t = 1, ..., T$), the first-order of the Long-DINA model can be expressed as

$$\text{logit}\,(P(Y_{nit} = 1 \mid \boldsymbol{\alpha}_{nt})) = \lambda_{i0t} + \lambda_{i(K)t} \prod_{k=1}^{K} \alpha_{nkt}^{q_{ikt}} + s_{im}\gamma_{nm}\,, \tag{3}$$

where $Y_{nit}$ denotes the response of person $n$ to item $i$ at occasion $t$; $\boldsymbol{\alpha}_{nt} = (\alpha_{n1t},\ldots,\alpha_{nKt})'$ denotes person $n$'s attribute profile on occasion $t$; $\lambda_{i0t}$ and $\lambda_{i(K)t}$ are the intercept and $K$-way interaction effect parameter for item $i$ on occasion $t$, respectively; $q_{ikt}$ is the element in a $I \times K$ Q-matrix on occasion $t$. $\gamma_{nm} \sim N(0, 1)$ be the $m$-th ($m = 1, \ldots, M$) specific dimension parameter for person $n$, and $\gamma_n$s are independent of each other. To simplify the computation, the item slopes on the $m$-th specific dimension are constrained to be equal as $s_{im} = s_m$ (Cai, 2010; Paek et al., 2014; Wang et al., 2016). Note that Equation 3 was a complete version of the first-order model, after specifying partial or all specific dimensions to be zero, some restricted models as illustrated in the empirical example would result.

***Second-order***. In the IRT framework, the multidimensional IRT models allow for the modeling of individual growth (te Marvelde, Glas, Van Landeghem, & Van Damme, 2006). Andersen (1985) proposed a between-item multidimensional Rasch model to measure individual differences on different occasions. Embretson (1991) proposed a within-item multidimensional Rasch model for learning and change. As the between-item multidimensionality is a special case of the within-item multidimensionality, Embreton's model can be taken as an extension of Andersen's model (Adams, Wilson, & Wang, 1997; von Davier et al., 2011). In addition, two- and three- parameter logistic multidimensional IRT models (e.g., von Davier et al., 2011; Paek et al., 2014) can also be employed in longitudinal studies.

In this study, a two-parameter logistic multidimensional higher-order latent structural model was used. For a given occasion $t$, the second-order of the Long-DINA model can be expressed as

$$\text{logit}(P(\alpha_{nkt} = 1 \mid \boldsymbol{\theta}_n)) = \delta_{kt}\theta_{nt} + \beta_{kt},\ \boldsymbol{\theta}_n = (\theta_{n1},\ldots,\theta_{nT})'\,, \tag{4}$$

where $\theta_{nt}$ is person $n$'s general ability on occasion $t$, $\delta_{kt}$ and $\beta_{kt}$ are the slope and intercept parameters of attribute $k$ on occasion $t$, respectively. $\theta_n s$ are constrained to be independent with $\gamma_n s$. Equation 4 is a between-attribute multidimensional model which is similar to Andersen's model. However, the major difference between these two models is that $\alpha_{nkt}$ in Equation 4 is latent but the item response in Andersen's model is observed. As a starting and a reference point for subsequent occasions, $\theta_{n1}$ is constrained to follow a standard normal distribution, $\theta_{n1} \sim N(0, 1)$, the mean values and variances of $\theta_{nt}$ ($t \geq 2$) are free to estimate. In addition, the same attributes are assumed to be measured on different occasions with the same latent construct on different occasions (Bianconcini, 2012), i.e., $K_t = K$. Correspondingly, the slope and intercept parameters of the $k^{\text{th}}$ attribute are constrained to be constants across occasions, $\delta_{kt} = \delta_k$ and $\beta_{kt} = \beta_k$. Each respondent's general ability and attribute mastery probabilities are allowed to change over occasions.

***Third-order.*** The most straightforward and general method assumes multiple general abilities follow a $T$-way multivariate normal distribution. Thus, the third-order of the Long-DINA model assumes that

$$\boldsymbol{\theta}_n = (\theta_{n1}, \ldots, \theta_{nT})' \sim \text{MVN}_T(\boldsymbol{\mu}_\theta, \boldsymbol{\Sigma}_\theta), \tag{5}$$

with a mean vector $\boldsymbol{\mu}_\theta = (\mu_1, \ldots, \mu_T)'$ and a variance and covariance matrix

$$\boldsymbol{\Sigma}_\theta = \begin{bmatrix} \sigma_1^2 & & \\ \vdots & \ddots & \\ \sigma_{1T} & \cdots & \sigma_T^2 \end{bmatrix},$$

where $\mu_1 = 0$ and $\sigma_1^2 = 1$; $\sigma_{1T}$ is the covariance of the first and $T^{\text{th}}$ general abilities. Additionally, the latent growth model-based strategy can also be employed in the third-order. That is, $\theta_{nt}$ is assumed to be a linear or nonlinear combination of the random coefficients or growth factors (Kohli & Harring, 2013) on occasions. Note that the latent growth model-based strategy is not employed in this study and can be one of the future explorations.

## Rebuilt the Longitudinal Data and Longitudinal Q-matrix

In the Long-DINA model, response data from different occasions were combined and calibrated simultaneously. Then, the longitudinal data is a $N \times \sum_{t=1}^{T} I_t$ matrix, and the longitudinal Q-matrix is constructed as a $\sum_{t=1}^{T} I_t \times TK$ matrix

$$\text{Longitudinal } \mathbf{Q} = \begin{bmatrix} \mathbf{Q}_1 & & & & \\ \vdots & \ddots & & & \\ \mathbf{0} & \cdots & \mathbf{Q}_t & & \\ \vdots & \ddots & \vdots & \ddots & \\ \mathbf{0} & \cdots & \mathbf{0} & \cdots & \mathbf{Q}_T \end{bmatrix}, \tag{6}$$

where $Q_t$ is the sub-Q-matrix for the test on the $t^{\text{th}}$ occasion. In such a case, the length of the estimated attribute pattern of each person was $TK$, which represented the attribute mastery status of $K$ attributes at $T$ occasions rather than $TK$ attributes for each person. Correspondingly, the posterior mixing proportions were computed at each occasion separately. Further, items from different occasions should be sequentially recoded for simultaneous estimation: items from $t^{\text{th}}$ occasion are recoded as $\sum_{t=0}^{t-1} I_t + 1$ to $\sum_{t=0}^{t} I_t$, where $I_0 = 0$.

## Overall and Individual Growth

Equations 3 to 6 together are the Long-DINA model. Using the Long-DINA model, both the overall and individual growth can be computed. The overall mean growth at the population level is $\hat{\mu}_{t+1} - \hat{\mu}_t$, and the overall scale change at the population level is $\hat{\sigma}_{t+1} / \hat{\sigma}_t$ (Paek et al., 2014). In the meantime, this model can also estimate the change in the mixing proportion of possible attribute patterns, the change of mean mastery probability of each attribute across all students, and the change of the number of students who master each attribute. For individual growth, the growth in the general ability can be computed as $\hat{\theta}_{n(t+1)} - \hat{\theta}_{nt}$, and changes in each attribute mastery status also can be reported.

In the Long-DINA model, the number of estimated parameters is $2\sum_{t=1}^{T}(I_t - d_t) + 2K + T(T-1)/2 + 2(T-1) + 3M$. More specifically, there are (1) $3M + 2\sum_{t=1}^{T}(I_t - d_t)$ item parameters including $2M$ parameters for anchor items, $2\sum_{t=1}^{T}(I_t - d_t)$ parameters for non-anchor items, and $M$ item slopes of special latent variables for anchor items; where $M$ is the total number of anchor items, $d_t$ is the number of anchor items on occasion $t$; (2) $2K$ latent structural parameters including $K$ attribute slopes and $K$ attribute intercepts; and (3) $T(T-1)/2 + 2(T-1)$ parameters of general abilities including $(T-1)$ averages, $(T-1)$ variances, and $T(T-1)/2$ covariances. Obviously, the number of model parameters is mainly influenced by the number of occasions, $T$. In addition, the complexity of model structure might also increase with the increase of $T$. Furthermore, the number of possible attribute patterns is $2^{KT}$, and it increases exponentially with $K$ and $T$. Therefore, the computational burden could be heavy when the number of occasions is large or even medium, which should be considered when applying the proposed model to real data.

Overall, as the proposed modeling approach is similar to the longitudinal IRT modeling approach, the interpretation of the proposed model is more straightforward than transition probability-based methods. The Long-DINA model use a multidimensional higher-order latent structural model to approximate the correlations among attributes at each occasion as well as across occasions. More importantly, local item dependence among a respondent's responses to the same anchor-items on multiple occasions can be modeled in the Long-DINA model. Essentially, the Long-DINA model can be seen as a special application of the higher-order, hierarchical DCM (Hansen et al., 2016) in longitudinal studies. The relationship between these two models is quite similar to that between the multidimensional IRT models and the longitudinal IRT models (e.g., te Marvelde et al., 2006).

## A Simulation Study

### Design and Data Generate

A simulation study was conducted to evaluate the parameter recovery of the Long-DINA model on different conditions. Three independent variables were manipulated including: (a) the sample sizes ($N$) at two levels of 200 and 500; (b) the qualities of anchor-items ($QA$) at two levels of high ($\lambda_{i0t} = -2.197$ and $\lambda_{i(K)t} = 4.394$) and moderate ($\lambda_{i0t} = -1.387$ and $\lambda_{i(K)t} = 2.774$). For the high-quality anchor-items, the aberrant response (i.e., guessing and slipping) probabilities are approximately equal to 0.1, while for the moderate-quality anchor-items, the aberrant response probabilities are approximately equal to 0.2. In practice, it is not common to use low-quality items as anchor-items; and (c) the number of occasions ($T$) at two levels of two and three.

Within each occasion, three attributes ($K_t = 3$) were measured by 20 items ($I_t = 20$), and first four items are used as anchor-items. A condition ($T = 2$) of simulated test structure is presented in Figure 2 as an example. The simulated Q matrices are presented in Figure 3. Non-anchor-item parameters were fixed at $\lambda_{i0t} = -2.197$ and $\lambda_{i(K)t} = 4.394$. For the general abilities, the correlations among them were set as 0.9. The overall mean growths were set as 0.5, and the overall scale changes were set as 1.25. Three specific dimensions were assumed to follow a standard normal distribution, and the slopes of the specific dimension were set as 0.8. In sum, the true person parameters including $T$ general abilities and 4 specific dimensions were generated from a ($T + 4$)-way multivariate normal distribution as $\mathbf{\Theta} \sim \mathrm{MVN}_{(T+4)}(\mathbf{\mu}, \mathbf{\Sigma})$. On each occasion, the true attribute pattern for each person is generated according to Equation 4, the true attribute intercept parameters were $\mathbf{\beta} = (-1, 0, 1)'$, and the true attribute slope parameters were $\mathbf{\delta} = (1, 1.25, 1.5)'$.

### Estimation and Analysis

Response data from different occasions were combined and calibrated simultaneously. Thus,

items on occasion 2 were recoded as items 21 to 40, and items on occasion 3 were recoded as items 41 to 60, accordingly. Then, for the conditions of $T = 2$, the longitudinal data was a $N \times 40$ matrix, and the longitudinal Q-matrix was constructed as a $40 \times 6$ matrix; for the conditions of $T = 3$, the longitudinal data was a $N \times 60$ matrix, and the longitudinal Q-matrix was constructed as a $60 \times 9$ matrix.

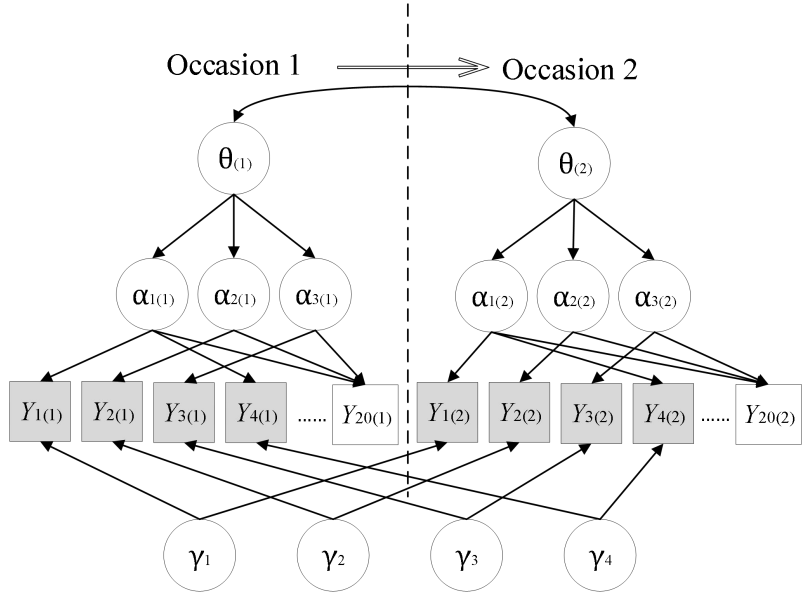

**Figure 2**. A condition ($T = 2$) of simulated test structure in simulation study.
*Note*. occasion in parenthesis.

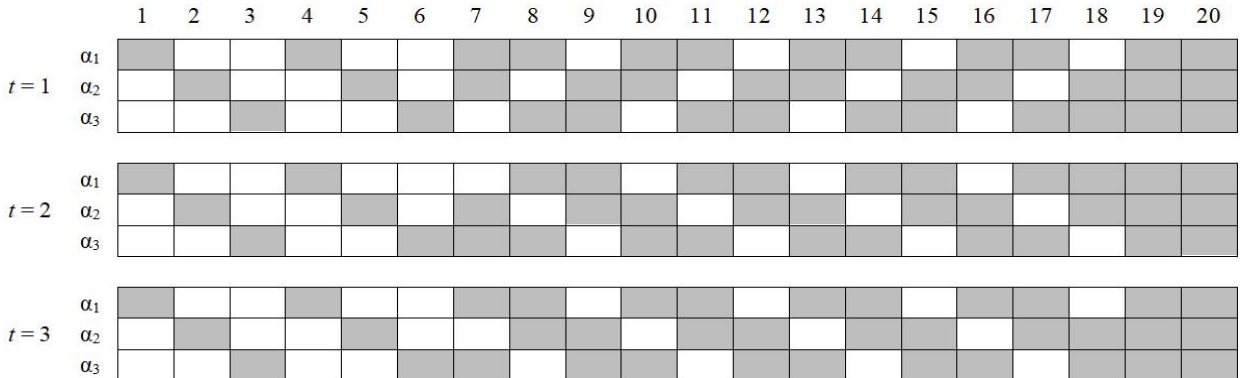

**Figure 3**. $K \times I_t$ Q' matrices of three occasions in the simulation study.
*Note*. $t = t$th occasion; gray = 1, blank = 0. When $T = 2$, only first two Q matrices were used.

For the model parameter estimation, flexMIRT® version 3.5 (Cai, 2017) was used. In flexMIRT®, the default Bock-Aitkin EM algorithm (Bock & Aitkin, 1981) was used for parameter estimation, and the Richardson extrapolation method was used to compute standard error. Specifically, the maximum number of cycles was set as 20,000 and 100 for the E-step and M-step, respectively; and the convergence criteria was $10^{-4}$ and $10^{-7}$ for the E-step and the M-step, respectively. Sample codes with comments are provided in the *Appendix*.

Thirty replications were implemented in each simulated condition. To evaluate model parameter recovery, bias and root mean square error (RMSE) were computed. The attribute correct classification rate (ACCR) and the pattern correct classification rate (PCCR) were computed to evaluate the classification accuracy of individual attributes and profiles. Additionally, the recovery of the overall mean and scale growths across different occasions were also computed.

**Results**

Figure 4 summaries the recovery of item parameters. First, the recovery of the intercept parameters was better than that of the interaction parameters. Further, the mean bias and mean RMSE for the study condition with a sample size of 200 were larger than those with a sample size of 500, indicating that a larger sample size led to better recovery of item parameters. In addition, the number of occasions and the quality of anchor-item had little effect on the recovery of item parameters.

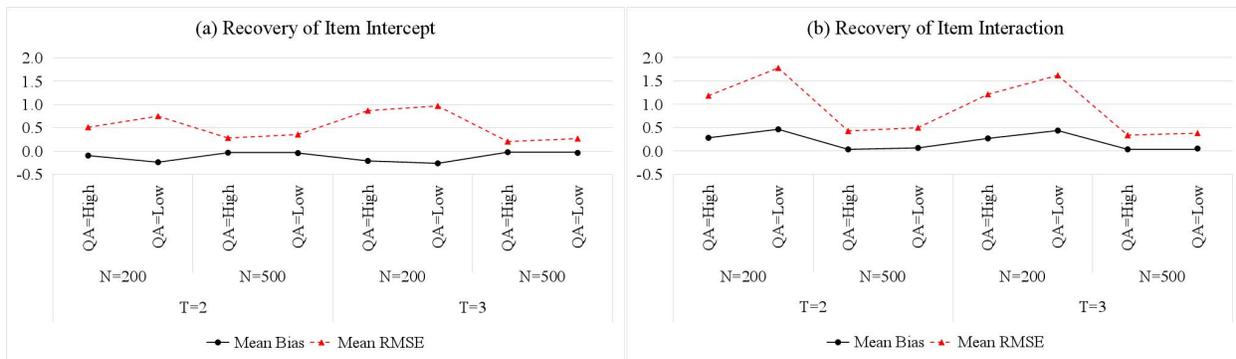

**Figure 4**. Recovery of item parameters.
*Note*, T = the number of occasions; N = sample size; QA = the quality of anchor item.

The recovery of attributes on different occasions is summarized in Table 1. The PCCR focuses on whether $K$ attributes can be correctly recovered on a given occasion; in contrast, the Longitudinal PCCR focuses on whether all $TK$ attributes can be correctly recovered. It can be found that the value of ACCR and PCCR both increased with time. According to the Longitudinal PCCR, it is evident that anchor-items with high quality improved the recovery of the attributes, and it is less evident that a larger sample size improved the recovery of the attributes. In addition, the Longitudinal PCCR decreased as the number of occasions increased.

**Table 1**. Recovery of Attributes in Different Simulation Condition.

| $T$ | $N$ | $QA$ | $t$ | ACCR | | | PCCR | Longitudinal PCCR |
|---|---|---|---|---|---|---|---|---|
| | | | | $\alpha_{1(t)}$ | $\alpha_{2(t)}$ | $\alpha_{3(t)}$ | | |
| 2 | 200 | High | 1 | 0.99 | 0.97 | 0.95 | 0.92 | 0.86 |
| | | | 2 | 0.99 | 0.97 | 0.96 | 0.93 | |
| | | Low | 1 | 0.97 | 0.96 | 0.95 | 0.89 | 0.83 |
| | | | 2 | 0.98 | 0.97 | 0.96 | 0.92 | |
| | 500 | High | 1 | 0.99 | 0.98 | 0.96 | 0.93 | 0.87 |
| | | | 2 | 0.99 | 0.98 | 0.97 | 0.94 | |
| | | Low | 1 | 0.97 | 0.97 | 0.95 | 0.89 | 0.83 |
| | | | 2 | 0.98 | 0.97 | 0.96 | 0.92 | |
| 3 | 200 | High | 1 | 0.99 | 0.98 | 0.96 | 0.93 | 0.85 |
| | | | 2 | 0.99 | 0.98 | 0.97 | 0.94 | |
| | | | 3 | 0.99 | 0.99 | 0.98 | 0.96 | |
| | | Low | 1 | 0.97 | 0.96 | 0.95 | 0.90 | 0.79 |
| | | | 2 | 0.98 | 0.97 | 0.96 | 0.92 | |
| | | | 3 | 0.98 | 0.98 | 0.98 | 0.95 | |
| | 500 | High | 1 | 0.99 | 0.98 | 0.96 | 0.93 | 0.85 |
| | | | 2 | 0.99 | 0.97 | 0.99 | 0.95 | |
| | | | 3 | 0.99 | 0.99 | 0.98 | 0.96 | |
| | | Low | 1 | 0.97 | 0.97 | 0.95 | 0.90 | 0.80 |
| | | | 2 | 0.98 | 0.97 | 0.96 | 0.92 | |
| | | | 3 | 0.98 | 0.98 | 0.97 | 0.95 | |

*Note*, $T$ = the number of occasions; $t$ = $t$-th occasion; $N$ = sample size; $QA$ = the quality of anchor-items; ACCR = attribute correct classification rate; PCCR = pattern correct classification rate.

Table 2 presents the recovery of the general abilities on different occasions. For occasion 1, virtually all conditions resulted in similar mean absolute bias; For occasions 2 and 3, the mean absolute bias was a little bit higher. Overall, the effects of the sample size and the quality of anchor-items were not evident on the recovery of the general abilities. Further, the RMSE of $\theta_{t+1}$ is larger than that of $\theta_t$, which means that the accuracy in the recovery of the general abilities diminished with time.

**Table 2**. Recovery of the General Abilities.

| $T$ | $N$ | $QA$ | $\theta_{(1)}$ | | $\theta_{(2)}$ | | $\theta_{(3)}$ | |
|---|---|---|---|---|---|---|---|---|
| | | | MA_Bias | M_RMSE | MA_Bias | M_RMSE | MA_Bias | M_RMSE |
| 2 | 200 | High | 0.34 | 0.62 | 0.35 | 0.72 | | |
| | | Low | 0.35 | 0.63 | 0.35 | 0.76 | | |
| | 500 | High | 0.35 | 0.63 | 0.37 | 0.71 | | |
| | | Low | 0.36 | 0.64 | 0.37 | 0.72 | | |
| 3 | 200 | High | 0.31 | 0.57 | 0.33 | 0.66 | 0.36 | 0.73 |
| | | Low | 0.32 | 0.58 | 0.34 | 0.67 | 0.41 | 0.77 |
| | 500 | High | 0.31 | 0.57 | 0.33 | 0.65 | 0.38 | 0.72 |
| | | Low | 0.31 | 0.57 | 0.33 | 0.65 | 0.39 | 0.72 |

*Note*, $T$ = the number of occasions; $N$ = sample size; $QA$ = the quality of anchor-item; $AI$ = the number of anchor-items. MA_Bias = mean absolute Bias across all respondents; M_RMSE = mean RMSE across all respondents.

Table 3 summarizes the recovery of the overall mean and the overall scale growth. For the overall mean growth, the bias is close to zero across all conditions; by contrast, for the overall scale growth, negative biases can be found, indicating that the Long-DINA model underestimated overall scale changes. Larger sample sizes seem to help the recovery, especially in terms of RMSE. The quality of anchor items did not evidently affect the recovery of these parameters.

**Table 3**. Recovery of the Overall Mean Growth and the Scale Growth.

| T | N | QA | Change of Occasion | Overall Mean Growth | | Overall Scale Growth | |
|---|---|---|---|---|---|---|---|
| | | | | bias | RMSE | bias | RMSE |
| 2 | 200 | High | $t_1 \rightarrow t_2$ | –0.02 | 0.15 | –0.09 | 0.29 |
| | | Low | $t_1 \rightarrow t_2$ | 0.03 | 0.19 | –0.03 | 0.21 |
| | 500 | High | $t_1 \rightarrow t_2$ | –0.02 | 0.08 | –0.11 | 0.18 |
| | | Low | $t_1 \rightarrow t_2$ | 0.03 | 0.11 | –0.10 | 0.17 |
| 3 | 200 | High | $t_1 \rightarrow t_2$ | –0.01 | 0.16 | –0.14 | 0.24 |
| | | | $t_2 \rightarrow t_3$ | 0.01 | 0.12 | –0.09 | 0.21 |
| | | Low | $t_1 \rightarrow t_2$ | –0.02 | 0.12 | –0.13 | 0.24 |
| | | | $t_2 \rightarrow t_3$ | 0.01 | 0.18 | –0.12 | 0.25 |
| | 500 | High | $t_1 \rightarrow t_2$ | –0.01 | 0.10 | –0.15 | 0.19 |
| | | | $t_2 \rightarrow t_3$ | –0.03 | 0.08 | –0.18 | 0.20 |
| | | Low | $t_1 \rightarrow t_2$ | –0.01 | 0.08 | –0.10 | 0.17 |
| | | | $t_2 \rightarrow t_3$ | –0.02 | 0.11 | –0.12 | 0.19 |

*Note*, $T$ = the number of occasions; $N$ = sample size; $QA$ = the quality of anchor-item; $AI$ = the number of anchor-items.

The recovery of attribute intercept and slope parameters are presented in Figure 5. For the attribute intercepts, the bias is close to zero across all conditions, while the bias for the attribute slopes is slightly larger. Large sample sizes seem to help the recovery, especially in terms of RMSE. On the contrary, the quality of the anchor items has no evident effect on the recovery of the attribute intercept parameters.

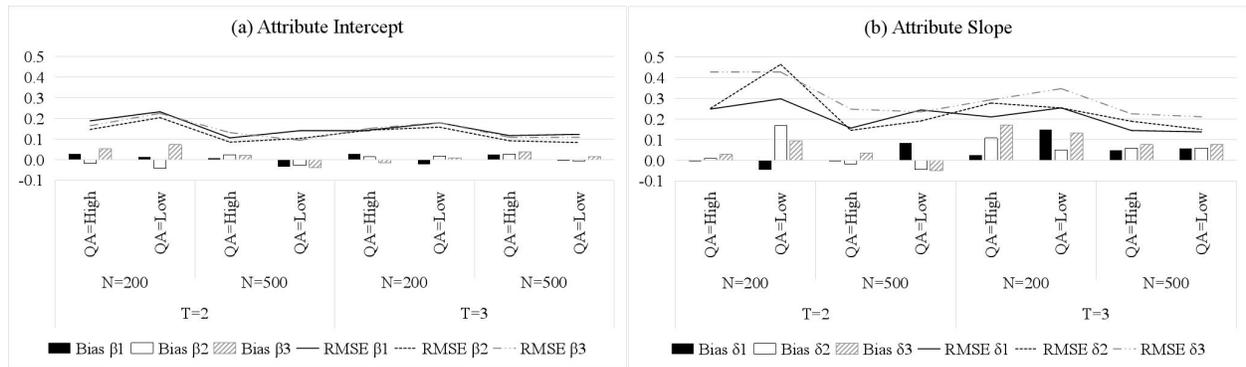

**Figure 5**. Recovery of the Attribute Intercept and Attribute Slope Parameters.
*Note*, T = the number of occasions; N = sample size; QA = the quality of anchor item.

## An Empirical Example

## Data description

Item response data from a physics achievement test about electric current and voltage were used to illustrate the application of the proposed Long-DINA model. Response data were available for three occasions. On occasion 1, 264 eighth grade students from 7 classrooms took part in the assessment in a school in Hangzhou, Zhejiang Province, China. After one week, 221 students from 6 classrooms remained on occasion 2. Another week later, 209 students from the same 6 classrooms remained on occasion 3. Among the 209 students, 7 students missed data collection on occasion 2. Thus, 202 respondents took part in all three tests. The same four attributes were assessed by all tests, namely, ($\alpha_1$) electric current; ($\alpha_2$) voltage; ($\alpha_3$) circuit analysis; and ($\alpha_4$) Ohm's law (resistance).

There were 17 items in the first two tests. Items 1 to 5 were dichotomous fill-in-the-blanks-items, items 6 to 15 were dichotomous multiple-choice items, and the last 2 constructed-response items were polytomously scored. Among the 20 items in the last test, items 1 to 6 were dichotomous fill-in-the-blanks-items, items 7 to 17 were dichotomous multiple-choice items, and last 3 constructed-response items were polytomously scored. For the current study, only dichotomous items were used. Items 1, 3, 6, 7, and 8 on occasion 1 were the same as items 2, 5, 9, 12, and 15 on occasion 2. Items 1 and 8 on occasion 1 were the same as items 5 and 16 on occasion 3, and items 7 and 10 on occasion 2 were the same as items 13 and 8 on occasion 3. Three Q matrices and test structure are presented in Figures 6 and 7, respectively. Students with missing responses to more than 7 items were removed while other missing data were treated as missing at random. The final cleaned data set contained 197 students, 15 dichotomous items in the first two occasions and 17 dichotomous items on the last occasion.

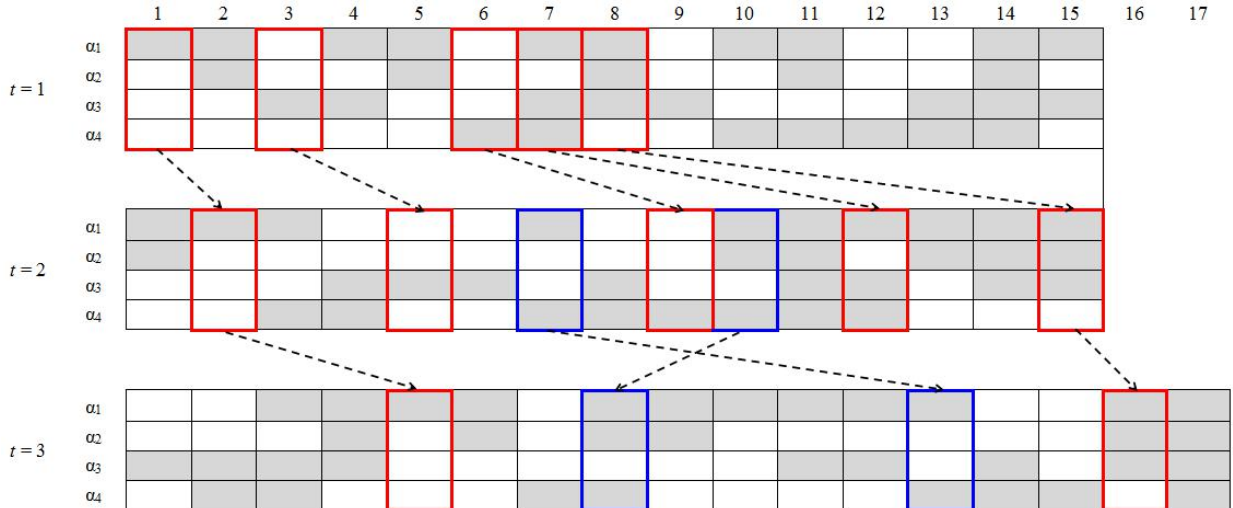

**Figure 6**. Three $K \times I_t$ Q' matrices for the empirical example where blank means "0", gray means "1", red or blue square represents anchor-items.

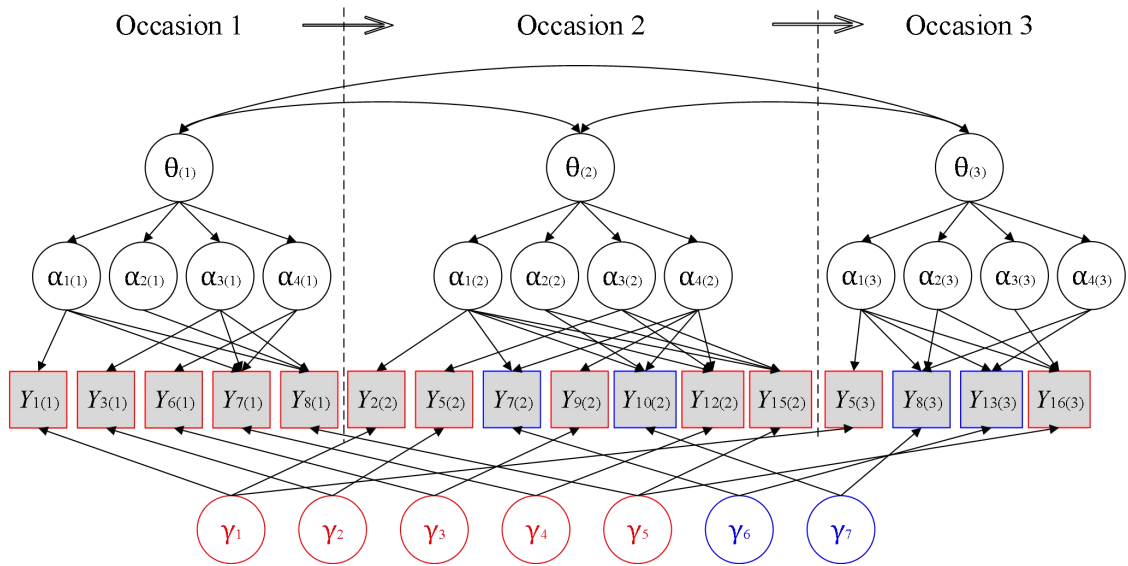

**Figure 7**. Test structure for the empirical example.
*Note*, non-anchor-items are omitted. Occasion in parenthesis.

## Analysis

Consistent with the simulation study, response data from different occasions were combined and calibrated simultaneously. Likewise, items on occasion 2 were recoded as items 16 to 30 and those on occasion 3 were recoded as items 31 to 47, accordingly. Thus, the longitudinal data was a 197 × 47 matrix, the longitudinal Q-matrix was constructed as a 47 × 12 matrix.

Two models were fitted to the data, a complete model (denoted as cLong-DINA), in which seven specific dimensions ($\gamma_1$ to $\gamma_7$) were included for all anchor-items; and a simple model (denoted as sLong-DINA) that ignored any specific dimensions. As aforementioned, $\theta_{n1}$ and all $\gamma_m$s were constrained to follow a standard normal distribution, and the item slopes on each $\gamma_m$ were constrained to be equal and need to be estimated. The $M_2$ statistic (Hansen et al., 2016) was used to evaluate the absolute model-data fit, and the Akaike's information criterion (AIC; Akaike, 1974) and Bayesian information criterion (BIC; Schwarz, 1978) were computed for each model to evaluate the relative model-data fit. The likelihood ratio test (i.e., $\Delta$ –2 log-likelihood, $\Delta$ –2LL) was also employed as the sLong-DINA model is nested within the cLong-DINA model.

## Results

Table 5 presents the model-data fit indexes of the compared two models. The value of $M_2$ for the cLong-DINA model was 1091.88, with 1030 degrees of freedom; and the RMSEA based on $M_2$ has a value of 0.02. By contrast, the value of $M_2$ for the sLong-DINA model was 1097.96, with 1037 degrees of freedom; and the RMSEA based on $M_2$ has a value of 0.02. Such results indicating both the cLong-DINA model and sLong-DINA model appear to provide reasonable good fit. Additionally, –2LL of cLong-DINA model is slightly better. This is expected because cLong-DINA model is more general than the sLong-DINA model. However, AIC and BIC both chose the sLong-DINA model as a better fitting model, and the likelihood ratio test also shows that the sLong-DINA model does not fit significantly worse than the cLong-DINA model ($\Delta$–2LL = 1.61, $df$ = 7, $p$ > 0.05). The estimated $s_m$ for each specific dimension is presented in Table 6. Under the cLong-DINA model, only estimates of $s_1$ and $s_3$ are higher than 0.01, which means that local item dependence among the anchor-items had limited impact. This may also explain why AIC and BIC tend to choose the sLong-DINA model. Thus, only the results pertain to the sLong-DINA model are discussed next.

**Table 5**. Summary of Model-Data Fit in the Empirical Example.

| Model | $M_2$ | $df$ | $p$ | RMSEA | NP | –2LL | AIC | BIC |
|-------|-------|------|-----|-------|-----|------|-----|-----|
| cLong | 1091.88 | 1030 | 0.088 | 0.02 | 98 | **9867.89** | 10063.89 | 10385.65 |
| sLong | 1097.96 | 1037 | 0.092 | 0.02 | 91 | 9869.50 | **10051.50** | **10350.27** |

*Note*. cLong = complete Long-DINA model; sLong = simple Long-DINA model; NP = number of estimated parameters; –2LL = –2 log-likelihood; AIC = Akaike's information criterion; BIC = Bayesian information criterion; *df* = degree of freedom; RMSEA = root mean square error of approximation.

**Table 6**. Estimated Item Slopes of Specific Dimensions (Standard Error in Parentheses).

| Fit Model | $s_1$ | $s_2$ | $s_3$ | $s_4$ | $s_5$ | $s_6$ | $s_7$ |
|-----------|-------|-------|-------|-------|-------|-------|-------|
| cLong | 0.74 | 0.00 | 0.47 | 0.00 | 0.00 | 0.01 | 0.00 |
|  | (0.40) | (1.10) | (0.74) | (1.21) | (0.46) | (0.33) | (0.70) |

*Note*. cLong = complete Long-DINA model.

Figure 8 presents the overall mean and scale growth of general ability with time. The overall means are $\hat{\mu}_2 - \hat{\mu}_1 = 0.34$ and $\hat{\mu}_3 - \hat{\mu}_2 = 0.42$; the overall mean growth from occasion 2 to 3 is a little larger than that from occasion 1 to 2. The overall scale growth is $\hat{\sigma}_2 / \hat{\sigma}_1 = 1.16$ and $\hat{\sigma}_3 / \hat{\sigma}_2 = 1.35$, which means that the gap between students becomes greater as time went by. To better understand these two concepts, we divide 197 students into (relatively) high and (relatively) low ability groups according to the median of the estimated general ability on occasion 1. Figure 9 presents the overall mean growth of such two groups with time. Obviously, the growth of the high-ability group is higher than that of the low-ability group, and the low-ability group grows a little and almost remains the same across occasions.

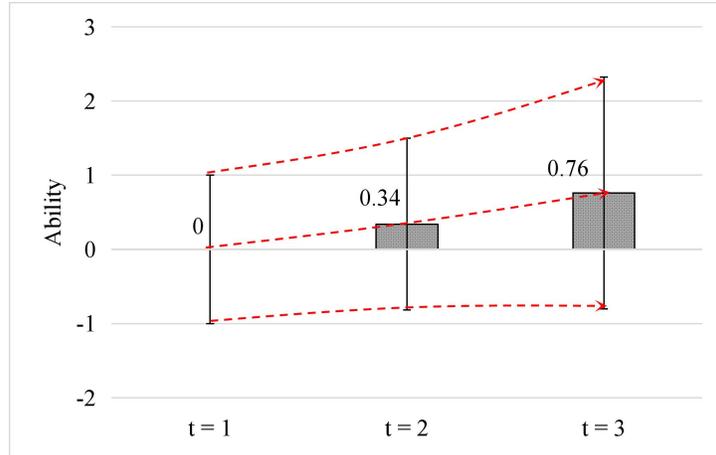

**Figure 8**. The overall mean and scale growth of the general ability with time.

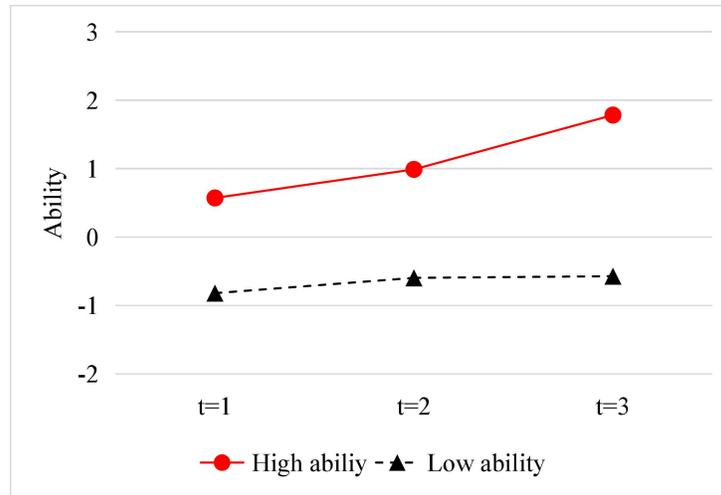

**Figure 9**. The overall mean growth of the high and low ability students.

Figure 10 presents the estimated overall growth of mean mastery probability across all students with time. In sum, the mean mastery possibilities of all four attributes increase with time. The mastery probability and the growth tendency of attribute 1 are close to those of attribute 2, and similar relationship can be found between attributes 3 and 4. Figure 11 presents the overall change of the number of students who mastered each attribute with time. Similarly, such numbers increase with time.

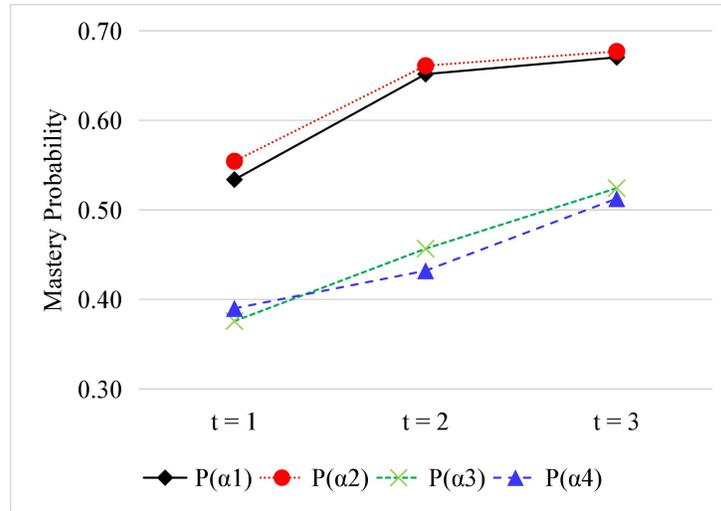

**Figure 10**. The overall growth of the mean mastery probability of each attribute with time.

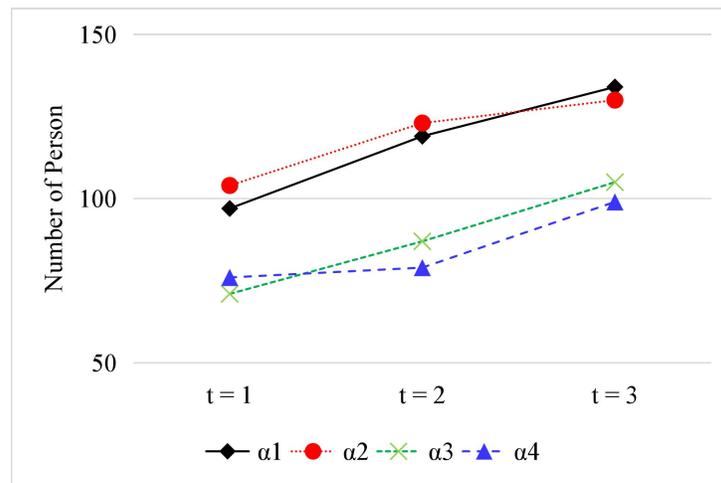

**Figure 11**. The overall growth of the number of students who mastered each attribute with time.

Table 7 presents the estimated means, variances, and covariances of the general abilities. The correlation between general abilities 1 and 2 is 0.95, between general abilities 1 and 3 is 0.91, and between general abilities 2 and 3 is 0.86. High correlations may be due to short time intervals. In addition, the model-implied (tetrachoric) correlations among attributes are presented in Table 8, which were computed after assigning a classification of all respondents. Moderate correlations are found among attributes regardless of the number of attributes within and across occasions. Such results indicate that the LTA-based method with attribute independence assumptions may over-simplify the real-world complexity.

**Table 7.** The Estimated Mean Vector and Variance and Covariance Matrix (Standard Error in Parentheses).

| | $\theta_{(1)}$ | $\theta_{(2)}$ | $\theta_{(3)}$ |
|---|---|---|---|
| $\theta_{(1)}$ | 1.00 | 0.95 | 0.91 |
| $\theta_{(2)}$ | 1.10 (0.21) | 1.34 (0.31) | 0.86 |
| $\theta_{(3)}$ | 1.43 (0.24) | 1.56 (0.36) | 2.44 (0.49) |
| $\mu_{(1)}$ | | $\mu_{(2)}$ | $\mu_{(3)}$ |
| 0.00 | | 0.34 (0.64) | 0.76 (0.46) |

*Note*, upper and lower triangular matrix is the covariances and correlations, respectively.

**Table 8.** The Model-implied Correlation Among Attributes.

| | $\alpha_{1(1)}$ | $\alpha_{2(1)}$ | $\alpha_{3(1)}$ | $\alpha_{4(1)}$ | $\alpha_{1(2)}$ | $\alpha_{2(2)}$ | $\alpha_{3(2)}$ | $\alpha_{4(2)}$ | $\alpha_{1(3)}$ | $\alpha_{2(3)}$ | $\alpha_{3(3)}$ | $\alpha_{4(3)}$ |
|---|---|---|---|---|---|---|---|---|---|---|---|---|
| $\alpha_{1(1)}$ | 1.00 | | | | | | | | | | | |
| $\alpha_{2(1)}$ | 0.42 | 1.00 | | | | | | | | | | |
| $\alpha_{3(1)}$ | 0.38 | 0.51 | 1.00 | | | | | | | | | |
| $\alpha_{4(1)}$ | 0.34 | 0.36 | 0.43 | 1.00 | | | | | | | | |
| $\alpha_{1(2)}$ | 0.42 | 0.58 | 0.56 | 0.27 | 1.00 | | | | | | | |
| $\alpha_{2(2)}$ | 0.41 | 0.65 | 0.52 | 0.50 | 0.69 | 1.00 | | | | | | |
| $\alpha_{3(2)}$ | 0.29 | 0.61 | 0.57 | 0.34 | 0.45 | 0.69 | 1.00 | | | | | |
| $\alpha_{4(2)}$ | 0.21 | 0.65 | 0.39 | 0.22 | 0.61 | 0.53 | 0.42 | 1.00 | | | | |
| $\alpha_{1(3)}$ | 0.38 | 0.39 | 0.36 | 0.22 | 0.19 | 0.10 | 0.28 | 0.12 | 1.00 | | | |
| $\alpha_{2(3)}$ | 0.20 | 0.63 | 0.49 | 0.40 | 0.38 | 0.59 | 0.57 | 0.35 | 0.45 | 1.00 | | |
| $\alpha_{3(3)}$ | 0.25 | 0.59 | 0.56 | 0.31 | 0.52 | 0.58 | 0.51 | 0.51 | 0.33 | 0.66 | 1.00 | |
| $\alpha_{4(3)}$ | 0.15 | 0.45 | 0.50 | 0.29 | 0.43 | 0.34 | 0.53 | 0.39 | 0.55 | 0.55 | 0.44 | 1.00 |

Figure 12 presents the change of the posterior mixing proportion with time. Take the (0000) and (1111) as two examples. The posterior mixing proportion of (0000) on occasion 1, 2, and 3 is 0.18, 0.15, and 0.12, respectively. In contrast, the posterior mixing proportion of (1111) at occasion 1, 2, and 3 is 0.14, 0.21, and 0.29. In sum, the proportion of students who master all attributes increases with time, and the proportion of students who master zero attributes decreases with time.

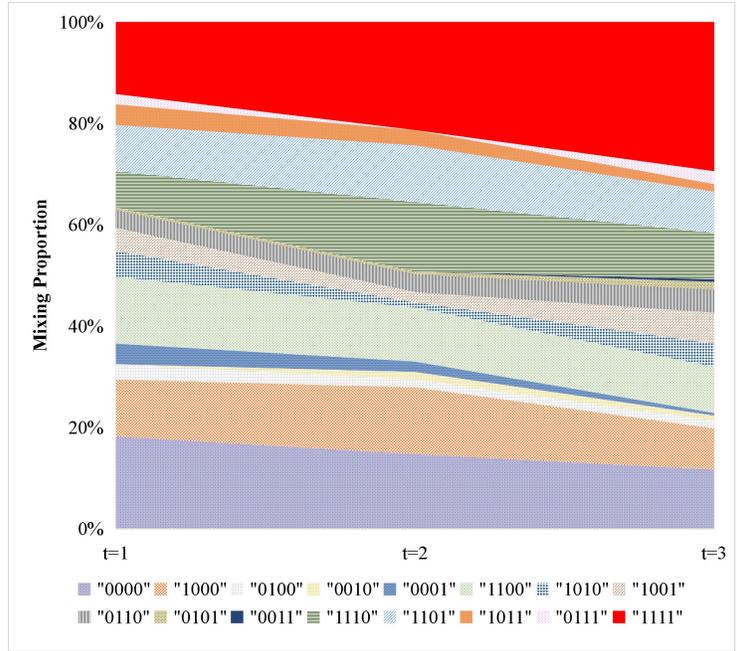

**Figure 12**. Overall change of posterior mixing proportion with time.

In addition to the overall growth, the growth of individuals also can be analyzed by the Long-DINA model. Three examples are presented in Table 9. For student ID = 2, after twice remedial teaching, the general ability increased significantly, from 0.37 to 1.28. Similarly, the attribute mastery status all change to 1. This indicates that the remedial teaching was effective for this person. By contrast, for student ID = 50, the general ability almost kept constant on three occasions. This means that the remedial teaching was not effective for this person. Similar conclusions can be drawn for this person's mastery of attributes. In addition, even though the general ability increased from 0.13 to 1.09, the student with an ID of 197 still has not mastered the fourth attributes after twice remedial teaching. Meanwhile, this student may have forgotten the second attribute during the second occasion.

Overall, the results from fitting the data to the Long-DINA model indicate that the remedial teaching was more effective for high-performing students than low-performing students. This result is consistent with the Matthew effect in education (Walberg & Tsai, 1983), which means students starting out at a higher level gain more on average than students starting at a lower level of proficiency (von Davier et al., 2011).

**Table 9**. Four Examples of Individual Growth of General Ability and Attributes with Time.

| Student ID | Growth | Parameter | $t = 1$ | $t = 2$ | $t = 3$ |
|---|---|---|---|---|---|
| | General Ability | $\theta$ | 0.37 | 0.74 | 1.28 |
| 2 | Attributes | $\alpha_1$ | 1 | 1 | 1 |
| | | $\alpha_2$ | 1 | 1 | 1 |
| | | $\alpha_3$ | 0 | 1 | 1 |
| | | $\alpha_4$ | 0 | 0 | 1 |
| | General Ability | $\theta$ | −1.04 | −0.94 | −0.90 |
| 50 | Attributes | $\alpha_1$ | 1 | 1 | 1 |
| | | $\alpha_2$ | 0 | 0 | 0 |
| | | $\alpha_3$ | 0 | 0 | 0 |
| | | $\alpha_4$ | 0 | 0 | 0 |
| | General Ability | $\theta$ | 0.13 | 0.63 | 1.09 |
| 197 | Attributes | $\alpha_1$ | 0 | 1 | 1 |
| | | $\alpha_2$ | 1 | 0 | 1 |
| | | $\alpha_3$ | 0 | 1 | 1 |
| | | $\alpha_4$ | 0 | 0 | 0 |

## Conclusions and Discussions

This study proposed a longitudinal diagnostic classification modeling approach for measuring individual growth, especially for the anchor-item design (also can be used in repeated measures design). Unlike the LTA-based method, the new modeling approach estimates the overall and individual growth and simultaneously retains the advantages of the higher-order latent structure (e.g., reduction in the number of model parameters) by constructing a multidimensional higher-order latent structure to take into account the correlations among multiple attributes. Additionally, potential local item dependence among anchor-items also can be taken into account. An empirical example was analyzed to illustrate the application and advantages of the proposed modeling approach.

The proposed modeling approach is the first attempt to measuring individual growth in cognitive diagnostic assessments by incorporating the multidimensional higher-order latent structure. Despite the promising findings, further study is still needed. For example, (a) only a DINA-based model was employed for illustrating the modeling approach, though the proposed modeling approach can be easily extended to the LCDM and its special cases. However, the performance of the proposed modeling approach based on other DCMs still need further investigation. (b) Currently, only the single group situation was considered, multiple group modeling (e.g., von Davier et al., 2011) can be extended in the future. (c) Additionally, in practice, students are nested in classrooms, and classrooms are further nested in schools. Thus, multilevel modeling (e.g., Fox & Glas, 2001; Huang, 2015; Jiao & Zhang, 2015) also can be incorporated into the third order of the proposed modeling approach. (d) Furthermore, theoretically polytomous attributes (Karelitz, 2004) provide more information than dichotomous attributes in describing the growth in longitudinal studies, as the former is more refined than the latter. Although the proposed modeling approach currently focuses on binary attributes, there is no conceptual challenge in extending the idea to model polytomous attributes by using the polytomous higher-order latent structural model (Zhan, Wang, Bian, & Li, 2016). (e) Detailed comparisons between other longitudinal diagnosis methods, e.g., transition probability-based methods, within the same conditions could be an interesting topic in the future. (f) In our empirical example, most respondents are allocated into the patterns that master the first or the second attribute; meanwhile, less respondents are allocated into the patterns that do not master the first two attributes (see Figure 12), which means these four attributes may follow a hierarchical structure (Leighton, Gierl, & Hunka, 2004). It is meaningful and practical to explore how to apply the Long-DINA model to hierarchical attribute situations. (g) Recently, some studies focus on utilizing response time information in cognitive diagnosis (e.g., Minchen, de la Torre, & Liu, 2017; Zhan, Jiao et al., 2018). How to incorporate response time information into the proposed longitudinal modeling approach is also an interesting topic (e.g., Wang, Zhang, Douglas, & Culpepper, 2018). It should be noted that the current Long-DINA model is complicated enough, which has already lead to heavy computing burdens, especially for complex test situations (e.g., more occasions, more attributes, and more anchor-items). Thus, the computing capability of computers should also be considered in further extension.

It is worthy of note that a necessary condition should be satisfied when using the proposed modeling approach, that is, the latent attributes measured by multiple tests must be invariant over time, i.e., the achievement construct does not shift across occasions. Occasionally, such assumption may be violated in practice. For instance, for cognitive areas (e.g., mathematics and reading), those target dimensions may change as students' grade levels increase (Wang & Jiao, 2009; Wang et al., 2013). In such cases, different attributes due to construct shift may be examined in multiple measures on different occasions. Therefore, the general abilities on different occasions may have different meanings (i.e., contain different target attributes). The complexity in computation and interpretation in this extension needs further exploration.

## Appendix: Sample flexMIRT® code for the Long-DINA model

```
<Project>
Title = "Sample flexMIRT code for the Long-DINA model";
Description = "Long-DINA model. Note that this sample code can be easily employed to the LCDM and its special cases";

<Options>
    Mode = Calibration;
    MaxE = 20000;
    MaxM = 100;
    Etol = 1e-4;
    Mtol = 1e-7;
    Quadrature = 49,6.0;
    Processors = 4;
    GOF = Extended;
    M2 = Full;
    SE = REM;
    SavePrm=Yes;
    SaveSco = Yes;
    Score = EAP;

<Groups>
%First_order%
    File = "C:\...\Score1.dat"; //locating the longitudinal data
    Varnames = v1-v40; // 20 items within each occasion
    Ncats(v1-v40) = 2; // dichotomous items; Can be easily extended to polytomous items
    Model(v1-v40) = Graded(2);
    Attributes = 6; // 3 attributes within each occasion
    Primary = 14; // 6 main effects and 8 interaction effects; Note that if ignore any specific dimension, pls remove this row.
    Dimensions = 17; // 14 primary dimensions and 3 specific dimensions; Note that if ignore any specific dimension, Dimensions = 14.
    Generate = (1,2),(1,3),(2,3),(1,2,3),(4,5),(4,6),(5,6),(4,5,6); //8 interaction effects

%Second_order%
    Varnames = a1-a6; //treat attributes in First_order as data in Second_order
    DM = First_order;
    Dimensions = 2; //two general abilities in Second_order

<Constraints>
//Attributes: included first 14 Dimensions in First_order.
//Be defined according to the longitudinal Q-matrix
Fix First_order,(v1-v40),MainEffect;
Free First_order,(v1,v4),MainEffect(1);
Free First_order,(v2,v5),MainEffect(2);
Free First_order,(v3,v6),MainEffect(3);
Free First_order,(v7,v10,v13,v16),Interaction(1,2);
Free First_order,(v8,v11,v14,v17),Interaction(1,3);
Free First_order,(v9,v12,v15,v18),Interaction(2,3);
Free First_order,(v19,v20),Interaction(1,2,3);
Free First_order,(v21,v24),MainEffect(4);
Free First_order,(v22,v25),MainEffect(5);
Free First_order,(v23,v26),MainEffect(6);
Free First_order,(v27,v30,v33,v36),Interaction(5,6);
Free First_order,(v28,v31,v34,v37),Interaction(4,6);
Free First_order,(v29,v32,v35,v38),Interaction(4,5);
Free First_order,(v39,v40),Interaction(4,5,6);

//Items 1-3 are anchor-items.
Equal First_order,(v1,v21), Intercept;
Equal First_order,(v2,v22), Intercept;
Equal First_order,(v3,v23), Intercept;
Equal First_order,(v1), MainEffect(1):First_order,(v21),MainEffect(4);
Equal First_order,(v2), MainEffect(2):First_order,(v22),MainEffect(5);
Equal First_order,(v3), MainEffect(3):First_order,(v23),MainEffect(6);
```

**Longitudinal Higher-Order Diagnostic Classification Modeling**

//Specific dimensions: the last 3 Dimensions in First_order.
Free First_order,(v1,v21),Slope(15);
Free First_order,(v2,v22),Slope(16);
Free First_order,(v3,v23),Slope(17);
Equal First_order,(v1,v21),Slope(15);
Equal First_order,(v2,v22),Slope(16);
Equal First_order,(v3,v23),Slope(17);

//2PL higher-order latent structural model (LSM)
Fix Second_order,(a1-a6), Slope;
Equal Second_order,(a1), Slope(1):Second_order,(a4), Slope(2);
Equal Second_order,(a2), Slope(1):Second_order,(a5), Slope(2);
Equal Second_order,(a3), Slope(1):Second_order,(a6), Slope(2);
Equal Second_order,(a1,a4),Intercept;
Equal Second_order,(a2,a5),Intercept;
Equal Second_order,(a3,a6),Intercept;

//Third-order describes the relationship between multiple general abilities in Second_order.
//The mean and variance of the first general ability are default fixed at 0 and 1, respectively.
Free Second_order, Mean(2);
Free Second_order, Cov(1,2);
Free Second_order, Cov(2,2);